\documentclass[11pt]{article}
\usepackage{latexsym}
\usepackage{amsmath,amsfonts,amsbsy,amssymb}
\usepackage[dvips]{graphicx,epsfig}

 \csname
@addtoreset\endcsname{equation}{section}

\def\mso{\mathfrak{so}}

\def\msl{\mathfrak{sl}}
\def\msp{\mathfrak{sp}}

\def\mhs{\mathfrak{hs}}

\def\mh{\mathfrak{h}}

\def\Real{{\mathbb R}}
\def\Comp{{\mathbb C}}
\def\integ{{\mathbb Z}}

\def\bec{\begin{center}}
\def\ec{\end{center}}
\def\a{\alpha} \def\ad{\dot{\a}} 

\def\b{\beta}  \def\bd{\dot{\b}} 
\def\c{\gamma} \def\cd{\dot{\c}}
\def\C{\Gamma}
\def\d{\delta} 

\def\e{\epsilon} 
\def\ve{\varepsilon}

\def\k{\kappa}
\def\vark{\varkappa}
\def\l{\lambda}

\def\m{\mu}
\def\n{\nu}

\def\s{\sigma}
\def\S{\Sigma}
\def\t{\tau}
\def\th{\theta} 
\def\Th{\Theta}

\def\O{\Omega}
\def\o{\omega}

\def\yb{{\bar y}}
\def\zb{{\bar z}}

\def\nn{\nonumber}
\newcommand{\eq}[1]{(\ref{#1})}

\def\be{\begin{equation}}
\def\ee{\end{equation}}
\def\bea{\begin{eqnarray}}
\def\eea{\end{eqnarray}}
\def\ba{\begin{array}}
\def\ea{\end{array}}

\def\ft#1#2{{\textstyle{{\scriptstyle #1}
\over {\scriptstyle #2}}}}

\def\ket#1{|#1\rangle}

\def\scs#1{\section{\bf{\sc \large #1}}}
\def\scss#1{\subsection{\bf{\sc  #1}}}

\def\ad{\dot\alpha}
\def\bd{\dot\beta}

\def\wV{\widehat{V}}

\def\wkappa{\widehat{\kappa}}

\def\widehatd{\widehat{d}}

\def\wf{\widehat{f}}
\def\wg{\widehat{g}}

\def\kb{\bar{k}}

\begin{document}



\vspace{8pt}

\begin{center}


{\Large\sc Biaxially Symmetric Solutions to 4D Higher-Spin Gravity}


\vspace{55pt}

C a r l o~~~~I a z e o l l a \\[15pt] 
{\it Dipartimento di Fisica, Universit\`a di Bologna \\
 and \\
 INFN, Sezione di Bologna\\
via Irnerio 46, I-40126 Bologna, Italy}\\[25pt]

P e r ~~~~ S u n d e l l\footnote{F.R.S.-FNRS Researcher with an Ulysse Incentive Grant for Mobility in Scientific Research.}\\[15pt]
{\it Service de M\'ecanique et Gravitation\\
Universit\'e de Mons\\
20, Place du Parc, 7000 Mons, Belgium }\\[25pt]


\vspace{10pt} {\sc\large Abstract}\end{center}

We review some aspects of biaxially symmetric solutions to Vasiliev's equations in four dimensional spacetime with negative cosmological constant.
The solutions, which activate bosonic fields of all spins, are constructed using gauge functions, projectors and deformed oscillators.
The deformation parameters, which are formally gauge invariant, are related to generalized electric and magnetic charges in asymptotic weak-field regions.
Alternatively, the solutions can be characterized in a dual fashion using zero-form charges which are higher-spin Casimir invariants built from combinations of curvatures and all their derivatives that are constant on shell and well-defined everywhere.

\vspace{1.3cm}

\noindent\emph{Invited contribution to the J. Phys. A special issue on ``Higher Spin Theories and Holography'' edited by M. Gaberdiel and M. A. Vasiliev.}

\vspace{0.5cm}

 \setcounter{page}{1}

\pagebreak

\tableofcontents

\vspace{2cm}


\scs{Introduction}\label{sec:intro}

Vasiliev's equations \cite{vasiliev,Vasiliev:2003ev} (see \cite{Vasiliev:1999ba,Vasiliev:1995dn,Bekaert:2005vh,Iazeolla:2008bp} for reviews) provide a classical description of a large class of higher-spin gravities: these are generally covariant gauge theories with (anti-)de Sitter vacua and perturbative spectra consisting of massless symmetric tensors and tensor-spinors of all possible ranks forming unirreps of underlying non-abelian higher-spin algebras. 
Vasiliev's framework for higher-spin gravities, which at present is the only known fully non-linear such framework, exhibits a level of complexity in between gravity and string field theory: the spectra and algebras are given by direct-product squares of singletons \cite{Flato:1978qz,vector}, that is, conformal fields on the spacetime boundary.
For suitable gauge algebras, field contents and couplings, these models are thus candidates for consistent truncations of string field theory in tensionless limit in the presence of a finite cosmological constant down to the first Regge trajectory, and by now there is indeed strong evidence that such models correspond holographically to weakly-coupled boundary theories, or rather, truncations thereof down to the sector of bilinear composites. 
Vasiliev's theory thus opens windows to the AdS/CFT correspondence, so far explored in both four \cite{Sezgin:2002rt,Sezgin:2003pt} and three bulk dimensions \cite{Campoleoni:2010zq,Gaberdiel:2012uj} (see more references in \cite{Gaberdiel:2012uj}), in the weak/weak coupling regimes, which has triggered the recent increase of interest in higher-spin gravity.

The aim of this brief review is to present some exact classical solutions that go beyond the aforementioned perturbative results.
The study of classical solutions facilitates the understanding of a number of interesting aspects of the theory: 
first, Vasiliev's equations are given in terms of locally defined fields living on charts, whereas a globally defined formulation must take into account a (choice of) structure group containing the transition functions used to glue together such field configurations across chart boundaries, as well as conditions on the fields at the boundary of the base manifold. Both of these data in general require a splitting of the one-form connection and its gauge parameters into a Yang-Mills-like subsector and a frame-like one containing a generalized soldering form \cite{Sezgin:2011hq}. Both of them can be used to construct observables, but of different geometric meaning. 
Second, given such geometric formulation, it can be used to construct super-selection sectors of the moduli space perturbatively as well as non-perturbatively \cite{us}.
Finally, as HS gravity exhibits rather special integrability properties, one may hope to actually be able to eventually provide an exact description of a classical moduli space with observables that can be (deformation-)quantized using geometric methods possibly along the lines of \cite{Boulanger:2011dd}.

In this work, we shall review some recently-found classical exact solutions \cite{us}, that possess at least two commuting Killing vectors, in the case of Vasiliev's twistor formulation in four dimensions; for exact solutions in lower dimensions, see \cite{Prokushkin:1998bq,Didenko:2006zd,Gutperle:2011kf}. In higher dimensions, the on-shell projection cannot be solved identically by means of the twistor transform. Thus, whether one works with twistor \cite{Sezgin:2001zs} or vector-oscillator \cite{Vasiliev:2003ev,vector,Bekaert:2005vh,Sagnotti:2005ns} formulations, there remain non-trivial internal constraints to be solved, and so far the only known exact solutions in dimensions higher than four are the maximally-symmetric spacetime vacua.

The plan of the paper is as follows: we continue by first presenting Vasiliev's equations, stressing their special features in the four-dimensional case. We then proceed discussing exact solutions.


\scs{Bosonic Vasiliev's equations in four dimensions}\label{sec:vaseqns}


\paragraph{\emph{Fundamental fields and kinematics}.} The basic variables of Vasiliev's formulation of higher-spin gravity are differential forms on $\mathfrak{C}$, a non-commutative symplectic manifold with symplectic structure $\O$, that we shall refer to as the correspondence space.
Locally, $\mathfrak{C}$ is the product of a phase-spacetime, containing the ordinary (commutative) spacetime, and internal directions, and it is endowed with an associative product $\star$ (containing the standard wedge-product among cotangent-space basis elements) and exterior derivative $\widehatd$ obeying
\bea 
\widehatd(\wf\star \wg)\ = \ (\widehatd\wf)\star \wg+ (-1)^{{\rm deg}(\wf)}\wf \star (\widehatd \wg)\ ,
\eea
where $\wf$ and $\wg$ are differential forms on $\mathfrak{C}$. The differential forms take their values in a unital associative algebra ${\cal A}$ whose product is assumed to be contained in $\star$ as well, \emph{i.e.} the differential forms are elements of the associative differential algebra\footnote{The elements of $\Omega({\mathfrak{C}})$ are thus composite operators given by functions of local non-commutative coordinates on $\mathfrak{C}$, which can be presented via their symbols comprising the expansion coefficients with respect to different bases, that correspond to choices of different ordering prescriptions; for further details, see for example Appendix B in \cite{us}.} ${\cal A}\otimes \Omega({\mathfrak C})$, assumed to admit a hermitian conjugation operation obeying
\bea  (\wf\star\wg)^\dagger \ = \ \wg^\dagger\star\wf^\dagger \ , \quad (\widehatd \wf)^\dagger \ = \  \widehatd(\wf)^\dagger \ ,  \eea

The field equations admit truncations to fields valued in the subalgebra $\mathbb{C}\subset {\cal A}$, that is, to elements of $\mathbb{C}\otimes \Omega({\mathfrak{C}})$, which is what we shall assume from now on. The correspondence space we shall be dealing with can be assumed to take the factorized form
\be \mathfrak{C}~\stackrel{\rm loc}{\cong}~ {\cal X}\times {\cal Y}\times {\cal Z}\ee
on local coordinate charts, where
\begin{itemize}

\item ${\cal Y}$ and ${\cal Z}$ are two $4$-dimensional real symplectic manifolds with coordinates $Y_{\underline{\a}}=(Y_{\underline{\a}})^\dagger$ and $Z_{\underline{\a}}=(Z_{\underline{\a}})^\dagger$ and with two-forms $\Omega_{\cal Y}=\frac{1}{2} dY^{\underline{\a}}dY^{\underline{\b}}C_{\underline{\a\b}}$ and $\Omega_{\cal Z}=-\frac{1}{2} dZ^{\underline{\a}}dZ^{\underline{\b}}C_{\underline{\a\b}}$, with non-trivial commutation relations normalized as follows: 
\bea [Y_{\underline{\a}},Y_{\underline{\b}}]_\star \ = \  2iC_{\underline{\a\b}}\ , \quad [Z_{\underline{\a}},Z_{\underline{\b}}]_\star \ = \ - 2iC_{\underline{\a\b}} \ , \quad   C_{\underline{\a\b}}\ := \ \left(\ba{cc}\e_{\a\b} & 0 \\ 0& \e_{\ad\bd}\ea\right)\ .\eea
The $\msp(4,\Real)$-quartets can be split into $\msl(2,\Comp)$-doublets as $Y^{\underline{\a}}=(y^\a,\yb^{\ad})$, with $\yb^{\ad}=(y^\a)^\dagger$, \emph{idem} for $Z^{\underline{\a}}$, with nontrivial commutation relations
\bea [y_\a,y_\b]_\star \ = \  2i\varepsilon_{\a\b}\ , \quad [z_\a,z_\b]_\star \ = \ - 2i\varepsilon_{\a\b} \ ,\eea
together with their hermitian conjugates. 

\item ${\cal X}$ is a spacetime manifold\footnote{This choice of $\mathfrak{C}$ corresponds to a truncation of a more general correspondence space locally admitting the factorization $\mathfrak{C}={\cal Y}\times {\cal Z}\times {\cal K}\times \check{\mathfrak{C}}$ where ${\cal K}$ is coordinatized by the outer kleinian elements $k$ and $\bar k$ obeying $ k^2=\kb^2=1, \ [k,\kb]_\star = 0, \ k^\dagger = \kb,\ \widehatd k=0  $
and having trivial commutators with all coordinates except for $\{k,y_\a\}_\star = 0 , \ \{k,z_\a\}_\star = 0 $ and their hermitian conjugates (see for example \cite{Vasiliev:1999ba}); and $\check{\mathfrak{C}}$ is a universal non-commutative manifold that one may take to be $T^\ast{\cal X}$, where ${\cal X}$ is a universal commutative manifold containing spacetime coordinatized with $x^\m$. The system on $T^\ast{\cal X}$ can be projected onto ${\cal X}$  (see \cite{us} for more details), which is what we assume in this paper.} coordinatized with $x^\m$.

\end{itemize}

The fundamental fields are a locally-defined zero-form $\widehat \Phi$; a locally-defined one-form $\widehat A$; and a globally-defined complex two form $(\widehat J,\,\,\widehat{\!\!\bar J})$. These master fields obey the reality conditions 
\be (\widehat \Phi,\widehat A,\widehat J,\widehat{\bar J})^\dagger~=~(\pi(\widehat \Phi),-\widehat A,-\widehat{\bar J},-\widehat J)\ .\label{reality}\ee
In bosonic models, they also obey the projections
\be \pi\bar\pi(\widehat \Phi,\widehat A)~=~(\widehat \Phi,\widehat A)\ ,\qquad \pi(\widehat J,\,\,\widehat{\!\!\bar J})~=~\bar\pi(\widehat J,\,\,\widehat{\!\!\bar J})~=~(\widehat J,\,\,\widehat{\!\!\bar J})\ ,\label{piofj}\ee
where $\pi$ and $\bar\pi$ are the involutive automorphisms defined by $\widehat d\,\pi=\pi\,\widehat d$, $\widehat d\,\bar\pi=\bar\pi\,\widehat d$ and
\be \pi(x^\m;y^\a,\yb^{\ad};z^\a,\zb^{\ad})~=~(x^\m;-y^\a,\yb^{\ad};-z^\a,\zb^{\ad})\ ,\qquad \pi(\widehat f\star\widehat g)~=~\pi(\widehat f)\star \pi(\widehat g)\ ,\ee
\be \bar\pi(x^\m;y^\a,\yb^{\ad};z^\a,\zb^{\ad})~=~(x^\m;y^\a,-\yb^{\ad};z^\a,-\zb^{\ad})\ ,\qquad \bar\pi(\widehat f\star\widehat g)~=~\bar\pi(\widehat f)\star \bar\pi(\widehat g)\ .\ee
In minimal bosonic models, the master fields obey the stronger projections
\be \tau(\widehat \Phi,\widehat A,\widehat J,\widehat{\bar J})~=~(\pi(\widehat \Phi),-\widehat A,-\widehat J,-\widehat{\bar J})\ ,\label{min}\ee
where $\tau$ is the graded anti-automorphism defined by $\widehat d\,\tau ~=~\tau\, \widehat d$ and
\be \tau(x^\m;Y^{\underline\a};Z^{\underline\a})~=~(x^\m;iY^{\underline\a};-iZ^{\underline\a})\ ,\qquad \tau(\widehat f\star\widehat g)~=~(-1)^{\widehat f\widehat g} \tau(\widehat g)\star \tau(\widehat f)\ ,\ee
and obeying $\tau^2=\pi\bar\pi$.
The perturbative spectra of the bosonic and minimal bosonic models consist of real Fronsdal fields of integer and even-integer spins, respectively, with each spin occurring in the spectrum with multiplicity one. 

The automorphisms $\pi$ and $\bar\pi$ are inner and generated via the adjoint action of inner Kleinian operators as follows:
\bea &\pi(\widehat f) \ = \   \widehat \kappa\star\widehat f\star\widehat \kappa\ ,\quad\qquad  \bar\pi(\widehat f)\ = \ \widehat{\bar{\kappa}}\star\widehat f\star\widehat{\bar{\kappa}}\ ,&\\[5pt]
&\wkappa  \ = \   \kappa_y\star\kappa_z \ ,\quad\qquad \widehat{\bar \kappa}  \ = \   \bar{\kappa}_{\bar y}\star\bar{\kappa}_{\bar z} \ ,&\label{fac}\eea
where $\k_y$ and $\k_z$ are the inner Kleinians \cite{Didenko:2009td}, obeying
\bea \k_y\star\k_y=1\ , \quad \{\k_y,y_\a\}_\star\ = \ 0 \ , \quad [\k_y,z_\a]_\star\ = \ 0 \ , \eea
\emph{idem}  for $\k_z$ upon exchanging $y$ and $z$ and for their hermitian conjugates. The inner Kleinians are given in Weyl order by
\bea  \k_y \ = \  2\pi \d^2(y) \ , \quad  \k_z \ = \  2\pi \d^2(z) \ ,\eea
\emph{idem} for $\bar \k_{\yb}$ and $\bar \k_{\zb}$ \cite{Didenko:2009td}  (for more details on Kleinians in various orderings, see Appendix B in \cite{us}).

An explicit, integral realization of the $\star$-product among functions of $(Y,Z)$ oscillators (corresponding to normal-ordering with respect to the creation/annihilation operators $\frac12( y_\a+z_\a,-i y_\a+i z_\a)$) is 
\be \widehat f_1\star\, \widehat f_2~=~ \int_{{\cal R}} {d^4U d^4V\over (2\pi)^4} e^{i( v^\a u_\a+\bar v^{\ad} \bar u_{\ad}) }\widehat f_1(y+u,\bar y+\bar u;z+u,\bar z-\bar u)\widehat f_2(y+v,\bar y+\bar v;z-v,\bar z+\bar v)\ ,\ee
where $(u,v)$ and $(\bar u,\bar v)$ are treated as real and independent variables.

The factorization property \eq{fac}, which holds in all orders, is crucial for the separation of twistor-space variables that we shall use in the solution Ans\"atze below. 

\paragraph{\emph{Field equations and deformed oscillators}.}The unfolded equations of motion of the four-dimensional bosonic higher-spin gravities that we shall study can be written as 
\be \O_{\cal Y}^{\star2}\star \widehat D\widehat \Phi~=~0\ ,\qquad  \O_{\cal Y}^{\star2}\star \left(\widehat F+{\cal F}(\widehat \Phi)\star \widehat J+\,\overline{\!\cal F}(\widehat \Phi)\star \,\,\widehat{\!\!\bar J}\,\right)~=~0\ ,\label{e1}\ee
\be \O_{\cal Y}^{\star2} \star \widehat d\,\widehat J~=~0\ ,\qquad \O_{\cal Y}^{\star 2} \star \widehat d \,\,\widehat{\!\!\bar J}~=~0\ ,\label{e3}\ee
with Yang-Mills-like curvatures $\widehat F:=\widehat d\,\widehat A+\widehat A\star\widehat A$ and $D\widehat \Phi:=\widehat d\,\,\widehat \Phi+[\widehat A,\widehat\Phi]_\pi$, where
$[\widehat f,\widehat g]_\pi:=\widehat f\star\widehat g-(-1)^{{\rm deg}(\widehat f){\rm deg}(\widehat g)}\widehat g\star\pi(\widehat f)$ for $\widehat f,\widehat g\in\O(\mathfrak{C})$.
The interaction ambiguities ${\cal F}$ and $\overline{\!\cal F}=({\cal F})^\dagger$ are given by
\be {\cal F}(\widehat \Phi)~=~\sum_{n=0}^\infty f_{2n+1}(\widehat \Phi\star\pi(\widehat\Phi)) \left(\widehat\Phi\star\pi(\widehat\Phi)\right)^{\star n}\star \widehat \Phi\ ,\ee
where $f_{2n+1}$ are complex-valued zero-form charges obeying 
\be \widehat d f_{2n+1}~=~0\ ,\label{f2n+1constraints}\ee
as we shall describe in more detail below.
Integrability requires the algebraic constraints
\be \widehat J\star \pi(\widehat \Phi,\widehat A)~=~(\widehat\Phi,\widehat A)\star \widehat J\ ,\qquad
\,\,\widehat{\!\! \bar J}\star \pi(\widehat \Phi,\widehat A)~=~(\widehat\Phi,\widehat A)\star \,\,\widehat{\!\! \bar J}\ ,\label{Jconstraints}\ee
modulo terms that are annihilated by $\O_{\cal Y}^{\star2}\star$. In other words, Eqs. \eq{e1}--\eq{e3} and Eq. \eq{f2n+1constraints} are compatible with $\widehat d^2\equiv 0$ modulo Eq. \eq{Jconstraints}, hence defining a universal (\emph{i.e.} valid on any ${\cal X}$) quasi-free associative differential algebra. Factoring out perturbative redefinitions of $\widehat \Phi$, the ambiguity residing in ${\cal F}$ reduces down to \cite{Vasiliev:1999ba,Sezgin:2011hq} 
\be {\cal F}~=~{\cal B}\star\widehat\Phi\ ,\qquad {\cal B}~=~\exp_\star\left(i\theta[\widehat\Phi\star\pi(\widehat\Phi)]\right)\ ,\label{calB}\ee\be \theta[\widehat\Phi\star\pi(\widehat\Phi)]~=~ \sum_{n=0}^\infty \theta_{2n}[\widehat\Phi\star\pi(\widehat\Phi)]\,\left(\widehat\Phi\star\pi(\widehat\Phi)\right)^{\star n}\ ,\label{Theta}\ee
which breaks parity except in the following two cases \cite{Sezgin:2003pt}:
\bea \mbox{Type A model (scalar)}&:& \theta~=~0\ ,\qquad P(\widehat \Phi,\widehat A,\widehat J)~=~(\widehat \Phi,\widehat A,\widehat J)\ ,\\[5pt]
\mbox{Type B model (pseudo-scalar)}&:& \theta~=~\frac{\pi}2\ ,\qquad P(\widehat \Phi,\widehat A,\widehat J)~=~(-\widehat \Phi,\widehat A,-\widehat J)\ ,\eea
where the parity operation is the automorphism of $\O(\mathfrak{C})$ defined by
\be P(x^\m,y^\a,\yb^{\ad},z^\a,\zb^{\ad})~=~(x^\m,\yb^{\ad},y^\a,-\zb^{\ad},-z^\a)\ ,\qquad \widehat  d P~=~P \widehat d\ .\ee
The gauge transformations read
\be \delta_{\widehat\e}\widehat \Phi~=~-[\widehat \e,\widehat\Phi]_\pi\ ,\qquad \delta_{\widehat\e}\widehat A~=~\widehat D\widehat\e\ ,\quad \delta_{\widehat\e}\widehat J~=~0\ ,\ee
with $\widehat D\widehat\e:=\widehat d\widehat\e+[\widehat A,\widehat\e]_\star$, and where $\widehat \e$ is subject to the same kinematic conditions as $\widehat A$. In globally-defined formulations (see \cite{us,Sezgin:2011hq} for more details), transition functions $T_I^{I'}$ (where the indices $I,I'$ denote charts and are understood in most of this paper) glue together the locally-defined configurations $(\widehat \Phi_I,\widehat A_I,\widehat J_I)$ as follows:
\be \widehat \Phi_I~=~(\widehat T_I^{I'})^{-1}\star\widehat \Phi_{I'}\star \pi(\widehat T_I^{I'})\ ,\qquad 
\widehat A_I~=~(\widehat T_I^{I'})^{-1}\star(\widehat A_{I'}+\widehat d)\star \widehat T_I^{I'}\ ,\qquad \widehat J_I~=~\widehat J_{I'}\ .\label{glue}\ee

The projection implied by the $\star$-multiplication by $\O_{\cal Y}^{\star2}$ can be solved locally on $\mathfrak{C}$ by taking the master fields to be forms on ${\cal X}\times {\cal Z}$ valued in the algebra $\O^{[0]}({\cal Y})$ of zero-forms on ${\cal Y}$. Thus 
\be \widehat A~=~\widehat U+\widehat V\ ,\ee
where $\widehat U= dx^\m \widehat U_\m(x;Z;Y)$, $\widehat V= dZ^{\underline\a}\widehat V_{\underline\a}(x;Z;Y)=dz^\a \widehat V_\a(x;Z;Y)+d\zb^{\ad} \widehat{\bar V}_{\ad}(x;Z;Y)$, 
and the algebraic constraints \eq{Jconstraints} admit the solution 
\be \widehat J~=~ -\frac{i}4 dz^\a \wedge dz_\a\,\widehat \k\ ,\qquad \,\,\widehat{\!\!\bar J}~=~-\frac{i}4 d\zb^{\ad}\wedge d\zb_{\ad}\,\widehat{\bar \kappa}\ .\label{J}\ee 
In order to find exact solutions, it is convenient to cast the remaining differential constraints into Vasiliev's original deformed-oscillator format:
\be d \widehat U+\widehat U\star \widehat U~=~0\ ,\qquad d\widehat \Phi+\widehat U\star\widehat \Phi-\widehat \Phi\star \pi(\widehat U)~=~0\ ,\label{MC}\ee
\be d\widehat S_{\underline\a}+[\widehat U,\widehat S_{\underline\a}]_\star~=~0\ ,\label{dSa}\ee
\be \widehat S_\a\star\widehat\Phi+\widehat\Phi\star\pi(\widehat S_\a)~=~0\ ,\quad
\widehat {\bar S}_{\ad}\star\widehat\Phi+\widehat\Phi\star\bar\pi(\widehat {\bar S}_{\ad})~=~ 0  \ ,\label{INT1}\ee\be
[\widehat S_\a,\widehat S_\b]_\star~=~ -2i\e_{\a\b}(1-{\cal B}\star\widehat \Phi\star\widehat \kappa)\ ,\quad [\widehat {\bar S}_{\ad},\widehat {\bar S}_{\bd}]_\star~=~ -2i\e_{\ad\bd}(1- \overline{\cal B}\star  \widehat\Phi\star\widehat {\bar \kappa}) \ ,\label{INT2}\ee\be
[\widehat S_\a,\widehat{\bar S}_{\ad}]_\star~=~ 0\ ,\label{INT3}\ee
where we have defined $d=dx^\m \partial_\m$ and 
\be \widehat S_{\underline\a}~=~Z_{\underline\a} -2i\widehat V_{\underline\a}~=~( \widehat S_{\a}, -\widehat{\bar S}_{\ad})~=~(z_\a-2i \widehat V_\a,-\zb_{\ad}+2i \widehat{\bar V}_{\ad})\ .\ee
The integrability of the system implies the gauge transformations 
\be \delta_{\widehat \e} \,\widehat \Phi~=~-[\widehat\e,\widehat\Phi]_\pi\ ,\quad \delta_{\widehat \e}\,\widehat S_{\underline\a}\ =\ -[\widehat \e,\widehat V_{\underline\a}]_\star\ ,\quad \delta_{\widehat \e}\, \widehat U~=~ d\widehat \e+[\widehat U,\widehat \e\,]_\star\ .\ee

\paragraph{\emph{Manifest Lorentz-invariance and component fields}.} Manifest local Lorentz covariance can be achieved by means of the field redefinition \cite{Vasiliev:1999ba,Sezgin:2002ru,Sezgin:2011hq}
\be \widehat W~:=~ \widehat U-\widehat K\ ,\qquad \widehat K~:=~ \frac1{4i} \left(\o^{\a\b} \widehat M_{\a\b}+\bar \o^{\ad\bd} \widehat {\overline M}_{\ad\bd}\right)\ ,\label{fieldredef}\ee
where $(\o^{\a\b},{\bar \o}^{\ad\bd})$ is the canonical Lorentz connection, and 
\bea \widehat M_{\a\b}&:=& \widehat M^{(0)}_{\a\b}+\widehat M^{(S)}_{\a\b}\ ,\qquad
\widehat {\overline M}_{\ad\bd}\: =\ \widehat{\overline M}^{(0)}_{\ad\bd}+\widehat{\overline{M}}^{(\bar S)}_{\ad\bd}\ ,\label{fullM}\eea
are the full Lorentz generators, consisting of the internal part
\bea \widehat M^{(0)}_{\a\b}&:=& y_{(\a} \star y_{\b)}-z_{(\a} \star z_{\b)}\ ,\qquad
\widehat{\overline M}^{(0)}_{\ad\bd}\ :=\ \yb_{(\ad}\star \yb_{\bd)}- \zb_{(\ad} \star \zb_{\bd)}\ ,\label{M(0)}\eea
rotating the $Y$ and $Z$ oscillators, and the external part
\bea \widehat M^{(S)}_{\a\b}&:=& \widehat S_{(\a}\star \widehat S_{\b)}\ ,\qquad
\widehat{\overline M}^{(\bar S)}_{\ad\bd}\ :=\  \widehat {\bar S}_{(\ad}\star \widehat {\bar S }_{\bd)}\ ,\label{M(S)}\eea
rotating the spinor indices carried by $(\widehat S_\a,\widehat{\bar S}_{\ad})$. As a result, the master equations read
\bea &\nabla\widehat W+\widehat W\star \widehat W + \frac1{4i} \left(r^{\a\b} \widehat M_{\a\b}+\bar r^{\ad\bd} \widehat {\overline M}_{\ad\bd}\right)\ =\ 0\ ,\quad \nabla\widehat \Phi+\widehat W\star\widehat \Phi-\widehat \Phi\star\pi(\widehat W)\ =\ 0\ ,\label{2.55}&\\[5pt]
&\nabla\widehat S_\a+\widehat W\star\widehat S_\a-\widehat S_\a\star \widehat W\ =\ 0\ ,\quad \nabla\widehat {\bar S}_{\ad}+\widehat W\star\widehat {\bar S}_{\ad}-\widehat {\bar S}_{\ad}\star \widehat W\ =\ 0&\\[5pt]
&
\widehat S_\a\star\widehat\Phi+\widehat\Phi\star\pi(\widehat S_\a)\ =\ 0\ ,\quad
\widehat {\bar S}_{\ad}\star\widehat\Phi+\widehat\Phi\star\bar\pi(\widehat {\bar S}_{\ad})\ =\ 0 &\\[5pt]
&
[\widehat S_\a,\widehat S_\b]_\star\ =\ -2i\e_{\a\b}(1-{\cal B}\star\widehat \Phi\star\widehat\kappa)\ ,\quad [\widehat {\bar S}_{\ad},\widehat {\bar S}_{\bd}]_\star\ =\ -2i\e_{\ad\bd}(1- \overline{\cal B}\star \widehat\Phi\star\widehat{\bar \kappa})&\\[5pt]
&
[\widehat S_\a,\widehat{\bar S}_{\ad}]_\star\ =\ 0\ ,&\label{2.56}\eea
where $r^{\a\b}:=d\o^{\a\b}+\o^{\a\c}\o^{\b}{}_\c$ and $\bar r^{\ad\bd}:=d\bar \o^{\ad\bd}+\o^{\ad\cd}\o^{\bd}{}_{\cd}$,
and
\bea \nabla \widehat W&:=& d\widehat W+\frac1{4i} \left[\o^{\a\b} \widehat M^{(0)}_{\a\b}+\bar \o^{\ad\bd} \widehat {\overline M}^{(0)}_{\ad\bd}~,~\widehat W\right]_\star\ ,\label{lorcovfirst}\\[5pt]
\nabla \widehat \Phi&:=& d\widehat \Phi+\frac1{4i} \left[\o^{\a\b} \widehat M^{(0)}_{\a\b}+\bar \o^{\ad\bd} \widehat {\overline M}^{(0)}_{\ad\bd}~,~\widehat \Phi\right]_\star\ ,\\[5pt]
\nabla \widehat S_\a&:=& d\widehat S_\a+\o_\a{}^\b\widehat S_\b+\frac1{4i} \left[\o^{\b\c} \widehat M^{(0)}_{\b\c}+\bar \o^{\bd\cd} \widehat {\overline M}^{(0)}_{\bd\cd}~,~\widehat S_\a\right]_\star\ ,\\[5pt]
\nabla \widehat S_{\ad}&:=& d\widehat S_{\ad}+\bar \o_{\ad}{}^{\bd}\widehat {\bar S}_{\bd}+\frac1{4i} \left[\o^{\b\c} \widehat M^{(0)}_{\b\c}+\bar \o^{\bd\cd} \widehat {\overline M}^{(0)}_{\bd\cd}~,~\widehat {\bar S}_{\ad}\right]_\star\ .\label{lorcovlast}\eea
Besides their manifest local Lorentz symmetry, these equations are by construction also left invariant under the local shift-symmetry with parameter $(\varsigma^{\a\b},{\bar \varsigma}^{\ad\bd})=dx^\m(\varsigma_\m{}^{\a\b},{\bar \varsigma}_\m{}^{\ad\bd})$ acting such that 
\be \delta_{\varsigma} (\widehat U,\widehat \Phi,\widehat S_\a,\widehat{\bar S}_{\ad})~=~0\ ,\qquad  \delta_{\varsigma} (\o^{\a\b},{\bar \o}^{\ad\bd})~=~ (\varsigma^{\a\b},{\bar \varsigma}^{\ad\bd})\quad\Rightarrow\quad \delta_{\varsigma}\widehat W~=~ -\frac1{4i} \left(\varsigma^{\a\b} \widehat M_{\a\b}+\bar \varsigma^{\ad\bd} \widehat {\bar M}_{\ad\bd}\right)\ .\ee
The canonical Lorentz connection can be embedded into the full theory by using the aforementioned shift-symmetry to impose 
\be\left.{\partial^2\over \partial y^\a \partial y^\b} \widehat W\right|_{Y=Z=0}~=~0\ ,\qquad \left.{\partial^2\over \partial \bar y^{\ad} \partial {\bar y}^{\bd}} \widehat W\right|_{Y=Z=0}~=~0\ .\label{shiftgauge}\ee

For the projection of Eqs. \eq{2.55}--\eq{2.56} to manifestly generally-covariant equations of motion for dynamical component fields in four-dimensional spacetime ${\cal X}_4$, see the Appendix D in \cite{us}. In essence, after choosing a manifestly $Sp(4;\Real)_{\rm diag}$-invariant ordering scheme, eliminating the auxiliary fields related to the unfolded description on ${\cal X}_4$ and ${\cal Z}$, and fixing suitable physical gauges (such as the \emph{universal twistor gauge condition} $z^\a\widehat{V}_\a=0$  and generalized holonomic gauges on $W_\m$), there remains a set of dynamical fields consisting of a physical scalar field 
\be \phi~\equiv ~C~:=~\widehat \Phi|_{Y=Z=0}\ ,\ee
which together with the self-dual Weyl tensors $C_{\a(2s)}$ ($s\geqslant 1$) make up the generating function ($s\geqslant 0$)
\be {\cal C}~:=~\widehat \Phi|_{Z=0,\yb =0}\ ,\qquad C_{\a(2s)}~:=~\left.\frac{\partial^{2s}}{\partial^{\a_1}\cdots\partial^{\a_{2s}}}{\cal C}\right|_{y=0}\ ,\label{2.60}\ee
and a tower of manifestly Lorentz-covariant, symmetric and doubly-traceless tensor gauge fields, or Fronsdal tensors, given by ($s\geqslant 1$)
\be \phi_{\mu(s)}~:=~\left.2i e^{\a_1\ad_1}_{(\mu_1} \cdots e^{\a_{s-1}\ad_{s-1}}_{\mu_{s-1}} \frac{\partial^{2s-2}}{\partial^{\a_1}\cdots\partial^{\a_{s-1}}\bar\partial^{\ad_1}\cdots \bar\partial^{\ad_{s-1}}} W_{\mu_s)}\right|_{Y=0}\ ,\ee
where $x^\mu$ are local coordinates on ${\cal X}_4$ and 
\be W~:=~ \widehat W|_{Z=0}~=~\left.\left(\widehat U-\frac1{4i} \left(\o^{\a\b} (y_\a\star y_\b+\widehat S_\a\star\widehat S_\b)+\bar \o^{\ad\bd} (\yb_{\ad}\star\yb_{\bd}+\widehat {\bar S}_{\ad}\star\widehat{\bar S}_{\bd})\right)\right)\right|_{Z=0}\ .\label{2.61}\ee 
%


\scs{Exact Solutions}\label{sec:exactsolns}



\scss{Gauge function method and moduli space}

Equations \eq{MC} and \eq{dSa} can be solved (on a chart $\mathfrak{C}_I$) by 
\be \widehat U_I~=~\widehat L^{-1}_I\star d\widehat L_I\ ,\quad \widehat \Phi_I~=~\widehat L^{-1}_I\star \widehat \Phi'\star \pi(\widehat L_I)\ ,\quad
\widehat S_{I;\underline\a}~=~ \widehat L^{-1}_I\star \widehat S'_{\underline\a}\star \widehat L_I\ ,\label{Lrot}\ee
where $\widehat L_I(X,Y,Z)$ is a gauge function, assumed to obey
\be \widehat L_I|_{X=Y=Z=0}~=~1\ ,\ee
and $(\widehat\Phi',\widehat S'_{\underline\a})$ are integration constants for the zero-forms on ${\cal X}_4$ given by 
\bea (\widehat\Phi',\widehat S'_{\underline\a})&=& (\widehat\Phi,\widehat S_{\underline\a})|_{X=0}\eea
and obeying the remaining twistor-space equations
\bea
&\widehat S'_\a\star\widehat\Phi'+\widehat\Phi'\star\pi(\widehat S'_\a)\ =\ 0\ ,\quad
\widehat {\bar S}{}'_{\ad}\star\widehat\Phi'+\widehat\Phi'\star\bar\pi(\widehat {\bar S}{}'_{\ad})\ =\ 0 &\label{prime1}\\[5pt]
&
[\widehat S'_\a,\widehat S'_\b]_\star\ =\ -2i\e_{\a\b}(1-{\cal B}\star\widehat \Phi'\star\kappa)\ ,\quad [\widehat {\bar S}{}'_{\ad},\widehat {\bar S}{}'_{\bd}]_\star\ =\ -2i\e_{\ad\bd}(1- \overline{\cal B}\star \widehat\Phi'\star\bar \kappa)&\label{prime2}\\[5pt]
&
[\widehat S'_\a,\widehat{\bar S}{}'_{\ad}]_\star\ =\ 0\ .&\label{prime3}\eea
Given a solution to these equations, the generating functions of the Weyl tensors and of the gauge fields respectively take the form ${\cal C}_I=\left(\widehat L^{-1}_I\star \widehat \Phi'\star \pi(\widehat L_I))\right|_{Z=0,\yb=0}$ and 
\be W_I~=~ \left.\widehat L^{-1}_I\star \left[d- \frac1{4i}\left(\o^{\a\b} \left( y_\a \star y_\b+ \widehat S'_\a\star\widehat S'_\b\right) + \bar\o^{\ad\bd} \left( \yb_{\ad} \star\yb_{\bd}+\widehat {\bar S}'_{\ad}\star\widehat {\bar S}'_{\bd}\right)\right)\right]\star \widehat L_I\right|_{Z=0}\ ,\label{Wmu}\ee
subject to \eq{shiftgauge}, which serves to determine $(\o_M^{\a\b},\bar \o_M^{\ad\bd})$.

In the following (omitting again the chart index $I$) we shall work with Gaussian gauge functions with factorized form
\be \widehat L(x|Y,Z)~=~L(x|Y)\star \tilde L (x|Z)\ ,\label{gaugef}\ee
realized as $\star$-exponentials of bilinears in $Y^{\underline\a}$ and $Z^{\underline\a}$, respectively. The $Y$-dependent factor reconstructs spacetime and will be chosen such that the flat connection $\Omega^{(0)}:=L^{-1}\star dL$ describes $AdS_4$.  One may choose $L$ to be manifestly Lorentz-covariant leading to \cite{Bolotin:1999fa,Sezgin:2005pv,Iazeolla:2008ix}
\be  L~=~\exp_\star (4i\xi  x^a  P_a) \ =\ {2h\over 1+h} \left[\exp {4ix^a P_a\over 1+h}\right]_{\rm Weyl}\ ,\qquad x^2  ~<~  1\ ,\qquad x^2~:=~ x^a x_a\ ,\label{L}\ee
\be\xi~:=~(1-h^2)^{-\ft12}\tanh^{-1}\sqrt{\ft{1-h}{1+h}}\ ,\qquad h~:=~\sqrt{1-x^2}\ . \ee
The corresponding vacuum connection $\O^{(0)}$ consists of the $AdS_4$ vierbein $e_{(0)}{}^{\a\ad}= - h^{-2}(\s^a)^{\a\ad}dx_a$ and Lorentz connection $\o_{(0)}{}^{\a\b}=- h^{-2} (\s^{ab})^{\a\b} dx_a x_b$ corresponding to presenting the metric in stereographic coordinates as\footnote{The metric remains well-defined for $x^2>1$ such that the regions $x^2<1$ and $x^2>1$ together yield a single cover of $AdS_4$.  For relations to global embedding coordinates and global spherically-symmetric coordinates, see Appendix A in \cite{us}.} $ds^2_{(0)}=4 (1-x^2)^{-2}dx^2$. On the other hand, allowing for a non-trivial $Z$-dependent factor fixes different gauges that may be helpful for studying the nature of certain singularities arising in some solutions that we shall review here. In order to illustrate how this issue arises, we shall begin with the trivial choice $\tilde L (x|Z)=1\!\!\!\!1$. 

A particular class of solutions, containing the exact solutions that we shall review here, admits perturbative expansions 
\be\widehat \Phi'~=~\sum_{n=1}^\infty \widehat \Phi^{\prime(n)}\ ,\qquad  \widehat S'_{\underline\a}~=~ \sum_{n=0}^\infty \widehat S^{\prime(n)}_{\underline\a}~\equiv~Z_{\underline\a}-2i\sum_{n=0}^\infty \widehat V^{\prime(n)}_{\underline\a} \ ,\label{solclass}\ee
where $(\widehat S^{\prime(n)}_{\underline\a},\widehat \Phi^{\prime(n)})$ are of the $n$th order in the integration constant  $\Phi'(Y)=\widehat \Phi'(Y,Z)|_{Z=0}$, and $\widehat S^{\prime(0)}_{\underline\a}$ is a flat connection in twistor space obeying $[\widehat S^{\prime(0)}_{\underline\a},\widehat S^{\prime(0)}_{\underline\b}]_\star = -2iC_{\underline{\a\b}}$.

Depending on the boundary conditions on $\widehat S^{\prime(0)}_{\underline\a}$ in twistor space there are various natural approaches to solving these equations: If the boundary conditions are chosen such that there exists a gauge where $\widehat V^{\prime(0)}_{\underline\a}=0$, one may adapt the perturbative scheme (see for example \cite{us}, Appendix D) to the case at hand. 
The solutions we shall discuss in this paper are of the form \eq{solclass} but are obtained by solving the deformed oscillator problem \eq{prime1}--\eq{prime3} using separation of variables in ${\cal Y}\times{\cal Z}$ space and the non-perturbative method of \cite{Prokushkin:1998bq,Sezgin:2005pv} adapted to the present case in \cite{us}. 
This method also encompasses non-trivial flat connections $\widehat V^{\prime(0)}_{\underline\a}$, essentially by activating Fock-space projectors in the space of functions on ${\cal Y}\times {\cal Z}$. The resulting solutions appear naturally in gauges that differ radically from the aforementioned universal twistor gauge in the sense that the space of residual symmetries is not isomorphic to $\mathfrak{hs}(4)$ or its non-minimal extension, as we shall discuss below. 

The space of solutions can thus be coordinatized by the following moduli (for a more detailed discussion on (iii) and (iv),  see \cite{Boulanger:2011dd,Sezgin:2011hq}):
\begin{itemize}
\item[(i)] local degrees of freedom contained in $\Phi'(Y)$\,;
\item[(ii)] boundary degrees of freedom contained in $\widehat L_I|_{\partial{\mathfrak C}}$ where $\partial{\mathfrak C}$ in particular contains the boundary of its four-dimensional spacetime sub-manifold\,; 
\item[(iii)] monodromies and projectors contained in flat connections $\widehat V^{\prime(0)}$ on ${\cal Z}\times {\cal Y}$ and $\widehat U^{(0)}$ on $T^\ast{\cal X}$ \,;
\item[(iv)] windings contained in the transition functions $\widehat T_I^{I'}$ between charts of the correspondence space\,.
\end{itemize}
In what follows, we shall mainly activate (i), (ii) and to some extent (iii), while we shall briefly discuss the possibility (iv) that more than one chart is required in reference to one particular family of solutions.


\scss{Solutions with spherical, cylindrical and biaxial symmetry}



Six infinite families of exact solutions admitting at least two commuting Killing vectors have been found in \cite{us} by extending the projector Ansatz used in \cite{Didenko:2009td} and combining it with the gauge function method. All of them can be obtained by solving the internal $Z$-space equations via the expansions
\bea & \widehat\Phi' \ = \ \sum_{\mathbf n} \nu_{\mathbf n} P_{\mathbf n}(Y)  \star \kappa_y \ ,&\label{3.6}\\[5pt]
& \widehat S_{\a}' \ = \ z_{\a}-2i \sum_{\mathbf n}  P_{\mathbf n} (Y) \star ( V_{\a})_{\mathbf n}(z)\ ,\qquad  \widehat{\bar S}_{\ad}'\ =\  \bar z_{\ad}-2i \sum_{\mathbf n}  P_{\mathbf n} (Y) \star (\bar V_{\ad})_{\mathbf n}(\bar z)\ ,& \label{3.6.1}\eea
where $\pi_z((V_{\a})_{\mathbf n})=-( V_{\a})_{\mathbf n}$,  $\bar\pi_{\bar z}((\bar V_{\ad})_{\mathbf n}) =  -(\bar V_{\ad})_{\mathbf n}$, $\nu_{\mathbf n}$ are \emph{a priori} complex constant deformation parameters and $P_{\mathbf n}=\pi\bar\pi(P_{\mathbf n})$ are projectors labelled by the (discrete) occupation numbers $\mathbf n:=(n_1,n_2)$ and assumed to obey 
\be P_{\mathbf n} \star P_{\mathbf m} \ = \ \delta_{\mathbf n,\mathbf m} P_{\mathbf n} \ . \label{3.31}\ee
and to form a set that is invariant under the operations 
$\pi$, $\dagger$ and $\tau$ and $\star$-multiplication by $ \kappa_y\bar \kappa_{\yb}$, such that 
\be \pi(P_{\mathbf n})~=:~P_{\pi(\mathbf n)}\ ,\qquad
(P_{\mathbf n})^\dagger~=:~P_{I(\mathbf n)}\ ,\qquad 
\tau(P_{\mathbf n})~=:~P_{\t(\mathbf n)}\ ,\label{indexmaps}\ee 
\be P_{\mathbf n}\star \kappa_y\bar \kappa_{\yb}~=:~ \kappa_{\mathbf{n}} P_{\mathbf n}\ ,\label{phasephi}\ee
with $\pi^2({\bf n})=I^2({\bf n})=\t^2({\bf n})=n$ and $(\kappa_{\mathbf{n}})^2=1$. The reality conditions fix the real or imaginary nature of the deformation parameters (which depends on $n$) \cite{us}.

This solution space\footnote{More generally, one can consider expansions over generalized, non-diagonal projectors $P_{\mathbf n|\mathbf n'}\sim \ket{\mathbf{n}}\langle \mathbf{n'}|$ (in this notation $P_{\mathbf n}:=P_{\mathbf n|\mathbf n}$). However, the latter lie along gauge orbits that can be reached from the diagonal solutions here considered \cite{us} (barring subtleties related to the admissibility of the corresponding gauge transformations). Moreover, while one may also allow for $Z$-dependent coefficients $\Phi_{\mathbf n}(Z)$ for the expansion of the Weyl zero-form \eq{3.6} and for non-holomorphic coefficients $( V_{\underline\a})_{\mathbf n}(Z)$ in \eq{3.6.1}, it is possible to show perturbatively in the initial datum $\n_{\mathbf n}:=\Phi_{\mathbf n}|_{Z=0}$ that one can always land on the form \eq{3.6}-\eq{3.6.1} via a partial gauge fixing (see \cite{us} for details).} forms an associative subalgebra of the $\star$-product algebra \cite{us}. Defining 
\be (\Sigma_{\underline\a})_{{\bf n}}~:=~Z_{\underline\a}-2i (V_{\underline\a})_{\mathbf n} \ , \qquad {\cal B}~\equiv~\exp_\star i\theta[\widehat \Phi\star\pi(\widehat \Phi)]~=:~\sum_{{\bf n}} P_{{\bf n}}\star  {\cal B}_{{\bf n}}\ ,\ee 
and using the factorization property \eq{fac}, one can show that: \emph{i)} the orthogonality of projectors splits the internal equations \eq{prime1}-\eq{prime3} into separate reduced deformed oscillator problems for every $\mathbf n$;  \emph{ii)} assuming $\nu_{\mathbf n}=\textrm{const}$ and $\pi_z(\Sigma_{\underline\a})_{{\bf n}}=-(\Sigma_{\underline\a})_{{\bf n}}$ solves \eq{prime1}
identically; and \emph{iii)} the holomorphicity of the reduced deformed oscillators $(\Sigma_{\underline\a})_{{\bf n}}$ solves \eq{prime3} identically. We are therefore left with 
\bea& [\S^{\mathbf{n}}_\a,\S^{\mathbf{n}}_\b]_\star\ =\ -2i\e_{\a\b}(1-\n_{\mathbf{n}}{\cal B}_{\bf n} \kappa_z)\ ,\label{deformed1}&\\[5pt]& [\bar \S^{\mathbf{n}}_{\ad},\bar \S^{\mathbf{n}}_{\bd}]_\star\ =\ -2i\e_{\ad\bd}(1-\kappa_{\bf n}\bar\n_{\mathbf{n}}\bar{\cal B}_{\bf n}\bar\kappa_{\zb})\ ,&\label{deformed2}\eea
which are defined modulo the residual holomorphic gauge transformations
\be \delta_{\e_{\mathbf{n}}}\Sigma^{\mathbf{n}}_\a~=~[\Sigma^{\mathbf{n}}_\a,\e^{\mathbf{n}}]_\star\ ,\qquad  \delta_{\bar\e_{\mathbf{n}}}\bar\Sigma^{\mathbf{n}}_{\ad}~=~[\bar\Sigma^{\mathbf{n}}_{\ad},\bar\e^{\mathbf{n}}]_\star\ ,\label{hologauge}\ee
\be \bar\partial_{\ad}\e^{\mathbf{n}}~=~0\ ,\qquad \partial_{\a}\bar \e^{\mathbf{n}}~=~0\ .\ee

\paragraph{\emph{Reduced deformed oscillators}.} These equations can be solved exactly by adapting the $\circ$-product method of \cite{Prokushkin:1998bq}, later refined in \cite{Sezgin:2005pv} (see also \cite{Iazeolla:2007wt}). A crucial difference with respect to the solutions found in those papers is that the deformation terms on the r.h.s. of \eq{deformed1}-\eq{deformed2} are distributional on ${\cal Z}$\footnote{Whether or not $\k_z$ is singular depends on the ordering prescription one is working with. Indeed, as shown in \cite{us}, while singular in Weyl ordering, $\k_z$ assumes a regular (gaussian) form in normal-ordering, and one can in fact solve for $(\S^{\mathbf n}_\a,\bar \S^{\mathbf{n}}_{\ad})$ in normal-ordering and then recast the so-obtained expressions of the deformed oscillators in Weyl ordering. The two procedures lead to the same result.}, admitting the limit representation $2\pi\d^2(z) = \lim_{\varepsilon\rightarrow 0}\frac{1}{\varepsilon}e^{-i\s\ft1\varepsilon w_z}~=~\s \k_z$, where $\s$ is a sign and a spin-frame $u^\pm_\a$ (in the terminology of \cite{PenroseRindler}, see also \cite{us}) has been introduced ($u^{\a+}u^-_\a=1$) in order to split $z^\pm:=u^{\pm\a}z_\a$, $w_z:=z^+z^-$, $[z^-,z^+]_\star = -2i$. Clearly, such splitting also breaks the manifest $SL(2;\Comp)$ covariance. Correspondingly, splitting 
\be \Sigma^{\mathbf{n}}_\a(z)~=~ u^-_\a \Sigma^{\mathbf{n}+}(z)-u^+_\a \Sigma^{\mathbf{n}-}(z)\ ,\qquad [\Sigma^{\mathbf{n}-},\Sigma^{\mathbf{n}+}] _\star~=-2i(1-{\cal B}_{\mathbf{n}}\nu_{\mathbf{n}} \k_z)\ ,\label{defoscpm}\ee
and representing $(\Sigma^{\mathbf{n}}_\a(z),\bar{\Sigma}^{\mathbf{n}}_{\ad}(\bar z))$ by the generalized Laplace transforms
\be \Sigma^{\mathbf{n}\pm} ~\equiv~ u^{\pm\,\a}\Sigma_\a^{\mathbf{n}} ~=~ 4z^{\pm}\int_{-1}^1 \frac{dt}{(t+1)^2}\,f^{\mathbf{n}\pm}_{\s_{\mathbf{n}}}(t)  \,e^{i\s_{\mathbf{n}}\ft{t-1}{t+1} w_z}\ ,\label{WSigmaansatz}\ee
where $(\s_{\mathbf{n}})^2=1$ can be chosen independently for each $\mathbf{n}$, one can show that \eq{deformed1} and \eq{deformed2}  are solved provided the transforms $f^{\mathbf{n}\pm}_{\s_\mathbf{n}}(t)$ obey the integral equations
\be  (f^{\mathbf{n}-}_{\s_{\mathbf{n}}}\circ f^{\mathbf{n}+}_{\s_{\mathbf{n}}})(t) ~=~ \d(t-1)-\frac{\s_{\mathbf{n}} {\cal B}_{\mathbf{n}}\n_{\mathbf{n}}}{2} \ ,\label{ringeq}\ee
where
\bea  (h_1 \circ h_2)(u) ~:=~ \int_{-1}^1 dt\int_{-1}^1 dt'\,h_1(t)\,h_2(t')\,\d(tt'-u)\ ,\label{ringo}\eea
defines a commutative and associative product on the space of functions on the unit interval. Note that, differently from the Lorentz-invariant case \cite{Sezgin:2005pv,Iazeolla:2007wt}, \eq{ringeq} constrains the $\circ$-product of two functions and not the square of a single one, and as a consequence its solution space is parameterized by an undetermined function. One can show however that that the latter is a gauge artefact (\emph{i.e.}, can be completely accounted for by the residual gauge symmetry \eq{hologauge}, see \cite{us}). One natural gauge choice is to work with symmetric solutions\footnote{Interestingly, the transformation to the most-asymmetric gauge choice, in which one sets, say, $f^{-}_\s=1$ and hence $f^{+}_\s=1-\frac{\s_{\mathbf{n}}{\cal B}_{\mathbf{n}}\n_{\mathbf{n}}}{2}$, also shows that in the latter the linear-in-$\n_{\mathbf{n}}$ correction in the deformed oscillators is actually exact (as it happens in the gauge used in \cite{Didenko:2009td}).} $f^{+}_\s= f^{-}_\s=f_\s $, and we shall therefore drop the $\pm$ referring to the spin-frame henceforth whenever not necessary. The solution in the holomorphic sector reads
\be \S_{\a}^{\mathbf{n}}~=~z_\a\left(1-2i \int_{-1}^1 \frac{dt}{(t+1)^2}\, j_{\mathbf{n}}(t) e^{i\s_{\mathbf{n}}\ft{t-1}{t+1} z^+ z^-}\right)\ , \label{Sigma}\ee
\be j_{\mathbf{n}}(t)~=~q_{\mathbf{n}}(t)-2\sum_{k=0}^\infty \th_{\mathbf{n},k} \left[1-{1+(-1)^k\over
2}\left(1-\sqrt{1-{\s_{\mathbf{n}}{\cal B}_{\mathbf{n}}\nu_{\mathbf{n}}\over 1+k}}\right)\right] p_k(t)\ ,\label{j}\ee
\be  q_{\mathbf{n}}(t)~=~-\frac{\s_{\mathbf{n}}{\cal B}_{\mathbf{n}}\nu_{\mathbf{n}}}{4}\,{}_1F_1\left[\frac{1}{2};2;\frac{\s_{\mathbf{n}}{\cal B}_{\mathbf{n}}\nu_{\mathbf{n}}}{2}\log\frac{1}{t^2}\right]\ ,\qquad p_k(t)~=~ {(-1)^k\over k!} \d^{(k)}(t)\ ,\ee
where $\s_{\mathbf{n}}\in\{\pm1\}$ and $\th_{\mathbf{n},k}\in\{0,1\}$. The second term in $j_{\mathbf{n}}(t)$ corresponds to the contribution of  $\circ$-product projectors $p_k\circ p_l=\d_{lk}p_k$ that activate a flat yet non-trivial part of the $Z$-space connection $V'_{\mathbf{n}\a}$ that survives in the limit $\nu_{\mathbf{n}}\to 0$ and receives $\nu_{\mathbf{n}}$-dependent corrections (see \cite{Iazeolla:2007wt,us} for more details). Such $Z$-space vacua are parameterized via the discrete parameters $\th_{\mathbf{n},k}$, that therefore represent true independent moduli along with the continuous deformation parameters $\nu_{\mathbf{n}}$. Note that, independently of the values of $\theta_{\mathbf{n},k}$, the branch-cut in \eq{j} can be chosen such that the internal connection is analytic for ${\rm Re}(\s_{\mathbf{n}}{\cal B}_{\mathbf{n}}\nu_{\mathbf{n}})<1$, where also the particular solution can be shown to be real analytic \cite{Sezgin:2005pv}. We shall only examine in the following the case where all $\theta_{\mathbf{n},k}=0$. Finally, we note that the singularity (both in the $t$-measure and in the exponent) in $t=-1$ of the reduced deformed oscillators \eq{Sigma} is precisely what reproduces the delta-function-like source $\k_z$ \cite{us} from their $\star$-commutator.

\paragraph{\emph{Projector algebras}.} It is possible to build the rank-one projectors $P_{n_1,n_2}$, on which the Ansatz \eq{3.6}-\eq{3.6.1} is based, starting from two commuting $\msp(4,\Comp)$ elements, which we shall denote $K^{(\pm)}=\ft18 K^{(\pm)}_{\underline{\a\b}}Y^{\underline\a}\star Y^{\underline\b}=\ft12(w_2\pm w_1)$, generating $\mh\cong \mso(2)_{(+)}\oplus \mso(2)_{(-)} \subset \msp(4;\Comp)$.  They obey ($n_i\in \integ+\ft12$)
\be P_{n_1,n_2}\star P_{n_1',n_2'} \ = \ \d_{n_1 n_1'}\d_{n_2 n_2'}P_{n_1,n_2} \ , \qquad (w_i-n_i) \star  P_{n_1,n_2} \ = \ 0 \ , \label{ortdiag}\ee
and admit the Weyl-ordered integral presentation ($\ve_i:=n_i/|n_i|$)
\bea P_{n_1 ,n_2} & = & 4(-1)^{\sum_i |n_i|-1}\,\oint_{C(\ve_1)} \frac{ds_1}{2\pi i} \frac{(s_1+1)^{n_1-\ft12}}{(s_1-1)^{n_1+\ft12}}\oint_{C(\ve_2)}\frac{ds_2}{2\pi i}\frac{(s_2+1)^{n_2-\ft12}}{(s_2-1)^{n_2+\ft12}}\,e^{-2\sum_i s_i w_i}\label{intproj} \\[5pt]
& = &4(-1)^{|n|-1}e^{-2(w_1+w_2)}L_{n_1-\ft12}(4w_1)L_{ n_2-\ft12}(4 w_2)\ , \qquad\label{nonintproj}\eea
where $C(\ve_i)$ are small contours encircling $\ve_i$. With the exception of the (anti-)ground-state projectors ($q=\ve_1\ve_2$) $P_{\ft{\ve_1}2,\ft{\ve_2}2}(K_{(q)})$,
which depend only on $K_{(+)}$ or only on $K_{(-)}$ and are therefore symmetric under the entire centralizer $\mathfrak{c}_{\msp(4,\Real)}(K_{(q)})$, these projectors depend on both $K_{(+)}$ and $K_{(-)}$ and are hence $\mh$-invariant. We therefore refer to the latter and to the solutions built on them as being biaxially symmetric (or axisymmetric). In order for the exponential in \eq{intproj} to give rise to a projector under $\star$-product it is crucial that the $\msp(4,\Comp)$ matrix $K^{(q)}_{\underline{\a\b}}$ satisfy
\be (K_{(q)})_{\underline{\a}}{}^{\underline\b}(K_{(q)})_{\underline{\b\c}}~=~-C_{\underline{\a\b}}\ .\label{K2=1a}\ee
This constraint leaves, as possible choices of two such commuting generators that are inequivalent up to $\msp(4,\Real)$ rotations, the following Cartan pairs: 
\be (E,J) \ , \qquad (J,iB) \ , \qquad (iB,iP) \ ,\ee
where $E:=P_0=M_{0'0}$ is the AdS energy, $J:=M_{12}$ is a spin, $B:= M_{03}$ is a boost and $P:=P_1=M_{0'1}$ is a translation\footnote{We denote with $M_{AB}$ ($A,B=0',0,1,2,3$) the $\mso(3,2)$ generators, that admit a realization as bilinears in $Y$ as $M_{AB}~=~ -\ft18  (\C_{AB})_{\underline{\a\b}}\,Y^{\underline\a}\star Y^{\underline\b}$ (where $(\C_{AB})_{\underline{\a\b}}$ are Dirac's gamma matrices) and can be split into Lorentz rotations $M_{ab}$ and $AdS$ translations $P_a$, $a,b=0,...,3$. We refer the reader to the Appendix A in \cite{us} for our $AdS_4$ and spinor conventions.}. Each pair gives rise to two families of solutions that are distinguished by the choice of the Cartan generator that the ground-state projector depends on, and which we will refer to in the following as \emph{principal} Cartan generator. The resulting six families can be therefore labelled as 
\be {\cal M}_{E}(E,J)\ ,\quad {\cal M}_{J}(E,J)\ ;\qquad {\cal M}_{J}(J,B)\ ,\quad {\cal M}_{iB}(J,B)\ ;\qquad {\cal M}_{iB}(B,P)\ ,\quad {\cal M}_{iP}(B,P)\ .\ee
When $K_{(+)}$ is principal, \emph{ i.e.} for the family ${\cal M}_{K_{(+)}}(K_{(+)},K_{(-)})$, the corresponding projectors have Cartan-eigenvalues such that $|K_{(+)}|>| K_{(-)}|$, while the opposite inequality $|K_{(+)}|<| K_{(-)}|$ holds when $K_{(-)}$ is principal. 

Each family contains a subset of solutions that possess the same symmetry under $\mathfrak{c}_{\msp(4,\Real)}(K_{(q)})$ of the ground-state, \emph{i.e.} one of the two $\mso(2)$ symmetries enhances to either $\mso(3)$ or $\mso(2,1)$. Such symmetry-enhanced solutions result from summing all axisymmetric projectors with fixed eigenvalue of the principal Cartan generator, in such a way that the dependence on the other Cartan generator drops out and one is left with the enhanced projectors ($n=\pm 1, \pm 2, ...$)
\bea {\cal P}_{n}(K_{(q)}) &=& \sum_{\tiny \ba {c}n_2+qn_1=n\\[-3pt] \e_1\e_2=q\ea}P_{n_1,n_2} \ = \ 4(-1)^{n-\ft{1+\ve}2} \,e^{-4K_{(q)}}L^{(1)}_{n-1}(8K_{(q)})\label{enhanced2}\\[5pt]& = & 2(-1)^{n-\ft{1+\ve}2}\,\oint_{C(\ve)} \frac{d\eta}{2\pi i}\,\left(\frac{\eta+1}{\eta-1}\right)^{n}\,e^{-4\eta K_{(q)}}\ ,\label{enhanced}\eea
that only depend on the principal Cartan generator.
 
Thus, for each family the Ansatz \eq{3.6} for the Weyl zero-form corresponds to choosing functions on ${\cal Y}$ that can be diagonalized over bases of eigenstates $\ket{\bf n}$ of the Cartan pairs (aside from the $\star$-multiplication with $\kappa_y$).
These solution spaces are coordinatized by deformation parameters $\nu_{\bf n}$ representing the eigenvalues of the Weyl zero-form master field in the aforementioned bases.
As we shall soon see, the principal Cartan generator not only determines the symmetry of the ground-state solution (and of the symmetry-enhanced subset of solutions), but also determines the spacetime behaviour of the Weyl tensors. 

It is interesting also to note that the ${\cal M}_{E}(E,J)$ family of solutions is based on projectors on scalar and spinor singleton states \cite{Iazeolla:2008ix}, \emph{i.e.} non-polynomial elements that are enveloping-algebra realizations of the states of boundary conformal scalar and spinor fields. 

While the integral and non-integral presentations of the projectors given in Eqs. \eq{intproj}-\eq{nonintproj} and \eq{enhanced2}-\eq{enhanced} are equivalent as long as $\star$-products among projectors with quantum numbers of the same sign are considered, the integral presentation ensures the orthogonality conditions in \eq{ortdiag} (simply via a change of variable, see Appendix F in \cite{us}) while preserving associativity also in the case that $\ve_1\ve_1'=-1$ or $\ve_2\ve_2'=-1$ whereas the non-integral one gives rise to divergencies. The integral presentation \eq{intproj} and \eq{enhanced2} is therefore required whenever reality or kinematical conditions (such as the minimal model truncation) force projectors with opposite quantum numbers in the expansions of the master-fields. 

\paragraph{\emph{Weyl zero form master-field}.}  Using the gauge function \eq{gaugef}, the $L$-rotation of Weyl zero-form master field in \eq{Lrot} gets the following form:
\be \widehat \Phi(x|Y,Z)~=~\sum_{\mathbf n }  \nu_{\mathbf{n}}P_{\mathbf{n}}^L(Y) \star\k_y\  \ ,\label{generalWeyl0f}\ee
where we use the notation $P_{\mathbf{n}}^L(Y)\equiv L^{-1}(x|Y)\star P_{\mathbf{n}}(Y)\star L(x|Y)$. The conjugation with $L$ induces an $x$-dependent rotation $(K^L_{(q)})_{\underline{\a\b}}(x) = L_{\underline\a}{}^{\underline\a'}(x) L_{\underline\b}{}^{\underline\b'}(x) (K_{(q)})_{\underline{\a'\b'}}$ of the Cartan matrices in the exponent of \eq{intproj} or \eq{enhanced} that preserves the conditions $[K^L_{(+)},K^L_{(-)}]_\star = 0$ and $(K^L_{(q)})^2_{\underline{\a\b}}=-C_{\underline{\a\b}}$. Each $(K^L_{(q)})_{\underline{\a\b}}(x)$ is a complexified $AdS_4$ global symmetry parameter satisfying $D^{(0)}  K^{L}_{\underline{\a\b}} = 0 $ and admitting the $SL(2,\Comp)$-decomposition \cite{Didenko:2009tc}
\bea K^L_{\underline{\a\b}} \ = \ \left(\ba{cc} \vark^L_{\a\b} & v^L_{\a\bd} \\[5pt] \bar{v}^L_{\ad\b} & \bar{\vark}^L_{\ad\bd} \ea \right) \ ,\qquad v^L_{\a\bd}~=~\bar v^L_{\bd\a}\ , \label{kv}\eea
yielding a complexified $AdS_4$ Killing vector $v^L_{\a\bd}(x)=\bar{v}^L_{\bd\a}(x)$ and the self-dual and anti-self-dual components $\vark^L_{\a\b}(x)$ and  $\bar{\vark}^L_{\ad\bd}(x)$, respectively, of the corresponding Killing two-form $\vark^L_{\m\n} : = \nabla^{(0)}_{\m} v^L_\n$.

Taking the $\star$-product with $\k_y$ and restricting our attention, for simplicity, to the symmetry-enhanced projectors\footnote{See \cite{us} for details of the Weyl zero-form in the axisymmetric case.} \eq{enhanced}, the resulting Weyl zero-form reads
\bea \widehat \Phi & = &  \frac{2}{\sqrt{(
\vark_{(q)}^L)^2}}\sum_{n=\pm 1,\pm2,...} (-1)^{n-\ft{1+\e}2}\nu_{n}\oint_{C(\e)}\frac{d\eta}{2\pi i\eta}\left(\frac{\eta+1}{\eta-1}\right)^{n}  \ \times\nn\\[5pt]&\times&\exp\left\{\frac{1}{\eta} \left[\ft12 y^\a
(\vark_{(q)}^L)^{-1}_{\a\b}y^\b+\ft12\yb^{\ad}(\bar{\vark}_{(q)}^L)^{-1}_{\ad\bd}\yb^{\bd}+iy^\a\yb^{\bd}(\vark_{(q)}^L)^{-1}_{\a\b}(v_{(q)}^L)^\b{}_{\bd}\right]\right\}
\label{enhancedPhi2} \ , \eea
where we denote (suppressing all the other labels) $\vark^2:=\ft12 \vark^{\a\b}\vark_{\a\b}$ and where $\vark^{-1}_{\a\b}=-\vark_{\a\b}/\vark^2$ and $\ve:=n/|n|$.  Note the dependence on the inverse square root of $(\vark_{(q)}^L)^2$ appearing in the prefactor and in the exponent (through $(\vark_{(q)}^L)^{-1}_{\a\b}$). For the solutions based on $\pi$-odd principal Cartan generators ($E$ and $iP$), for which the $x$-independent $K^{(q)}_{\underline{\a\b}}$ matrix is off-diagonal and $(\vark_{(q)})_{\a\b}=0$, the $x$-independent Weyl master zero-form $\widehat\Phi'$ has a delta-function-like behaviour in twistor space. The latter is thus softened by the spacetime dependence introduced via the gauge-function, and in particular $\sqrt{(\vark_{(q)}^L)^2}$ appears as the parameter of a limit representation of the delta function. From \eq{enhancedPhi2} it is also clear that the principal Cartan generator $K_{(q)}$ specifies, essentially through the determinant of its upper diagonal block $(\vark_{(q)}^L)^2$, the spacetime behaviour and the singularities of every spin-$s$ Weyl tensor. Note however that the singular behaviour of the individual Weyl tensors is not a higher-spin invariant statement, and that even a singular, delta-function-like behaviour of the master-field in some spacetime point can be an artefact of the chosen ordering prescription\footnote{See, for example, the comment in the Footnote 6.} (which is also a consequence of the fact that the fibre-space that the master-fields are valued in is infinite-dimensional). We shall later examine this issue at the level of certain higher-spin-invariant quantities. 

Let us now specialize \eq{enhancedPhi2} to the case of solutions based on the spherically-symmetric projectors ${\cal P}_n(E)$.  One can show that, starting from $K_{(+)}=E$ (\emph{i.e.}, $K^{(q)}_{\underline{\a\b}}=(\C_0)_{\underline{\a\b}}$), the rotation with the gauge function \eq{gaugef} gives rise to non-vanishing diagonal blocks with determinant $(\vark^L)^2=(\bar\vark^L)^2=-r^2$, where $r$ is the radial coordinate in the $AdS_4$ spherical coordinate system. The reality conditions on the master fields in this case require that $\n_n=i^n\m_n$, where $\m_n\in \Real$. For any fixed projector ${\cal P}_n(E)$, expanding in $y$ the Weyl zero-form and performing the auxiliary integration yields the physical  scalar ($s=0$) and an infinite tower of spherically-symmetric Type-D \cite{Petrov:2000bs,PenroseRindler} Weyl tensors of spin $s\geqslant 1$ of the form (up to real $n$-dependent numerical factors)
\bea C^{(n)}_{\a(2s)}\ \sim \ \frac{i^{n-1}\mu_n}{r^{s+1}}\,(\tilde u^+
\tilde u^- )^s_{\a(2s)} \ ,\label{nsWeylspher}\eea
where $(\tilde u^+_\a,
\tilde u^-_\a )$ are eigenspinors of $\vark^L_{\a\b}$ generating a spin-frame at every spacetime point where $(\vark^L)^2\neq 0$ (and are only dependent on the angular variables $(\th,\phi)$) \cite{us}. As first noted in the case $n=1$ in \cite{Didenko:2009td}, the spin-$2$ Weyl tensor coincides with that of an $AdS_4$-Schwarzschild black hole of mass $\m_n$, which here appears together with infinitely many partners of all integer spins. However, examining individual Weyl tensors only makes sense asymptotically (\emph{i.e.} for $r\to\infty$), where they are all small and as a result fields of different spin are weakly coupled. In strong-field regions, and in particular in the proximity of the apparent singularity in $r=0$, the higher-spin symmetry is fully realized and one should rather examine higher-spin invariant quantities. 

There are a few observations that one can make from \eq{nsWeylspher}. First, we note that the Weyl zero-form components are real for $n$ odd (\emph{i.e.} for solutions built on projectors  ${\cal P}_n(E)$ over combinations of states belonging to the scalar singleton representation) and imaginary for $n$ even (\emph{i.e.}, for solutions based on spinor singleton projectors). On the linearized higher-spin equations in the parity preserving case the gauge-field curvatures are equated to the above-derived Weyl zero-form components up to an extra factor $b=1$ (Type A model) or $b=i$ (Type B model). Therefore, one can in this sense regard the deformation parameters of the solutions based on the scalar singleton
as generalized electric charges (or generalized masses) in the Type A model and generalized magnetic charges (or generalized NUT charges) in the Type B model, and, conversely, those of the solutions based on the spinor
singleton as magnetic-like charges in the Type A model and electric-like charges in the Type B model, and the two models seem to be connected via a generalized electric/magnetic duality. 

Second, any solution based on a single projector is activated by a single deformation parameter that switches on the entire tower of Weyl tensors of all spins (and, in fact, interestingly enough the Didenko-Vasiliev solution, based on ${\cal P}_1(E)$ alone, appears to be extremal \cite{Didenko:2009td}). On the other hand, building the solutions over a generic expansion in projectors opens up the possibility of diagonalizing $\Phi$ with respect to the spin $s$ insted of the occupation number $n$, thus having different deformation parameters ${\cal M}_s$ in front of the spin-$s$ Weyl tensors that are sums of the $\n_n$'s with spin-dependent coefficients obtained from the contour integration,
\be {\cal M}_s \ = \ {\cal N}_{s}\sum_{n} (-1)^{n-\ft{1+\e}2}\,\n_n\oint_{C(\e)}\frac{d\eta}{2\pi i \eta^{s+1}}\left(\frac{\eta+1}{\eta-1}\right)^n \ ,\ee
 where ${\cal N}_{s}$ is an $s$-dependent normalization coefficient. 
  
While identifying ${\cal M}_s$ with a spin-$s$ asymptotic charge may be very tempting, it is complicated by the fact that these solutions have been obtained in a gauge in which the one-form field-strengths are non-linear in $\n_n$ \cite{us}: as a consequence, the Weyl 0-form components \eq{nsWeylspher} may differ from the linearized spin-$s$ gauge-field curvatures asymptotically by non-linear lower-spin constructs that may give contributions of the same order in the limit $r\to\infty$ and modify the proposed asymptotic charge by non-linear terms is the ${\cal M}_s$ parameters.
An additional complication arises from the fact that the gauge we work with is not the standard, universal twistor gauge of the perturbative analysis mentioned at the end of Section \ref{sec:vaseqns} , as we shall comment more on in the next section.  

Finally, specializing \eq{enhancedPhi2} to projectors ${\cal P}_n(J)$ that only depend on $K_{(-)}=J$ (i.e., $K^{(q)}_{\underline{\a\b}}=(\C_{12})_{\underline{\a\b}}$), one can check that the rotation with the gauge function \eq{gaugef} modifies the already non-vanishing determinant of the diagonal blocks as $(\vark^L)^2=(\bar\vark^L)^2=1+r^2\sin^2\th$, giving rise to a tower of Weyl 0-form components exhibiting cylindrical symmetry of the form 
\bea C^{(n)}_{\a(2s)} \ \sim \ \frac{i^{n+s+1}\m_n}{(1+r^2\sin^2\theta)^{\ft{s+1}2}}\,(\tilde u^+ \tilde u^-)^s_{\a(2s)} \ . \eea
Note that such Weyl tensors do not blow up anywhere and do not vanish at spatial infinity (they are constant along the $z$ axis, with a behaviour similar to that of the Melvin solution in General Relativity \cite{Melvin}). Moreover, since the Killing two-form is imaginary, for every fixed $n$ the electric/magnetic type of the type-D Weyl tensors flips according to whether the spin is even/odd, for $n$ odd, vicecersa for $n$ even.
These solutions are $\mso(2)_J\oplus\mso(2,1)_{E, M_{03},P_3}$-symmetric, and are built on the spacelike AdS Killing vector $\partial/\partial\varphi$ in the same way as the spherically-symmetric ones are based on the timelike vector $\partial/\partial t$, \emph{i.e.}, the $\mso(2,1)$ is the stability subalgebra of $\partial/\partial\varphi$. In other words, here the roles of $E$ and $J$ are exchanged, with respect to the rotationally-invariant case, and the corresponding solutions are based on projectors onto combination of states belonging to non-unitary analogues of the (anti-)supersingleton of fixed $J$ and vanishing energy.

\paragraph{\emph{Deformed oscillators and gauge fields}.} 
We now turn to examining the spacetime-dependent deformed oscillators. Defining $\widehat S^{\prime \pm}:=u^{\pm\a} \widehat S'_\a$, one has
\be \widehat S^\pm~:=~(L)^{-1}\star \widehat S^{\prime \pm}\star L~=~  z^\pm-2i\sum_{\mathbf{n}} P^L_{\mathbf{n}}\star  V_{\mathbf{n}}^\pm\ ,\label{Spm}\ee
where we recall that $V^{\mathbf{n}\pm}$ denotes the second term in \eq{Sigma}, which is a non-analytic function of the oscillators due to the singularity of both the measure and the exponent in $t=-1$. As explicitly shown in \cite{us}, the $\star$-product of the latter with the $L$-rotated, $x$-dependent projector $P^L_{\mathbf{n}}$ pushes such singular points outside the integration interval for generic values of $x$. In short, this happens because the $\star$-product with $P^L_{\mathbf{n}}$ gives rise to a Gaussian determinant depending on $\vark^L_{\a\b}$ that effectively shifts the potential singularities in the integral representation of $\widehat S^\pm_K$ to the zeroes of
\be (t+1)^2-i\s_{\mathbf{n}} (t^2-1)\vark^{L\a\b}(u^+_\a u^-_\b+u^-_\a u^+_\b)+(\vark^L)^2 (t-1)^2\nn\ee
\be =~ (t+1)^2+2i\s_{\mathbf{n}}\e_2 \Theta (t^2-1)\cos\th-\Theta^2(t-1)^2\nn\ee
\be~=\Big(\big(1+i\s_{\mathbf{n}}\e_2\Theta \big)\left(t+1\right)-2i\s_{\mathbf{n}}\e_2\Theta\Big)^2+2i\e_2\Th\s_{\mathbf{n}}(t^2-1)(\cos\th-1)\ ,\label{shift}\ee
and this same shift takes place at the denominator in the exponent\footnote{To be precise, this is the shift after the contour-integration (entering through the projector) has been carried out. If one keeps the contour integral, some of the terms in \eq{shift} will also contain powers of the auxiliary integration variables $(s_1,s_2)$ or $\eta$ appearing in \eq{intproj} and \eq{enhanced}. See \cite{us} for the details.}. Above, we have denoted $\Theta^2:=-(\vark^L)^2 $, and the factor of $\cos\th$ (where $\th$ is the polar angle in spherical coordinates) enters via the contraction of the $x$-dependent eigenspinors of $(\vark^L)_{\a}{}^\b$ with the rigid spin-frame $(u^+_\a, u^-_\a)$. The crux of the matter is that at any point $x$ in which $\Th$ is real and non-vanishing the singularity in the deformed oscillators acquires an imaginary part and is therefore pushed out of the integration domain, provided that $\th\neq \pi/2$! This is what happens, in particular, for the spherically-symmetric solutions. 

Recalling that the singularities in the deformed oscillators are inherited by the gauge fields via \eq{Wmu}, it is important to clarify at this point whether this singular behaviour at the equator is physical or not. However, the fact that it comes from a point-wise non-collinearity of two spin-frames suggests that it should be pure gauge. Indeed, one can modify the gauge function \eq{gaugef} as follows,
\be \widehat L_{(K)}(x|Y,Z)~=~L(x|Y)\star \tilde L_{(K)} (x|Z)\ ,\label{gaugefZ}\ee
with a non-trivial $Z$-dependent factor $\tilde L_{(K)}:~{\cal R}_{4}\rightarrow SL(2;\Comp)/C_{SL(2;\Comp)}(K^{L})$ (where $C_{SL(2;\Comp)}(K^{L})$ is the centralizer of the principal Cartan generator, that we shall here simply denote as $K$) that aligns the spin-frame of ${\cal Z}$ with the spin-frame of ${\cal Y}$ generated by the eigenspinors of $(\vark^L)_{\a}{}^\b$ (hence the labelling with the relevant global symmetry parameter $K$)\footnote{See Appendix E in \cite{us} for the detailed form of this change of twistor basis implemented by $\tilde L_{(K)}$.}. Substituting $L$ with $\widehat L_{(K)}$ in \eq{Spm}, one obtains 
\be \widehat S^\pm_{(K)}~:=~(\widehat L_{(K)})^{-1}\star \widehat S^{\prime \pm}\star \widehat L_{(K)}~=~ \tilde z^\pm_{(K)}-2i\sum_{\mathbf{n}} P^L_{\mathbf{n}}\star \tilde V^{\mathbf{n}\pm}_{(K)}\ ,\label{SKpm}\ee 
where now $\tilde z^\pm_{(K)}:=(\tilde L_{(K)})^{-1}\star z^\pm\star \tilde L_{(K)}=\tilde u^{\pm\a}_{(K)} z_\a$, where we denote with $(\tilde u^{+}_{\a},\tilde u^{-}_{\a})_{(K)}$ the $\vark^L$-eigenspinors, and
\be \tilde V^{\mathbf{n}\pm}_{(K)}~=~2i \tilde z^{\pm}_{(K)} \int_{-1}^1 \frac{dt}{(t+1)^2}\, j^\pm_{\mathbf{n}}(t) e^{i\s_{\mathbf{n}}\ft{t-1}{t+1} \tilde z^+_{(K)} \tilde z^-_{(K)}}\ .\label{intrep1}\ee
After one computes the $\star$-product of the latter with the projectors, the shift \eq{shift} takes place, but the alignment between the spin-frames now sets $\cos\th=1$. This has two consequences:

\begin{enumerate}

\item The deformed oscillators $\widehat S^\pm_{(K)}$ are real-analytic in $Y$ and $Z$
if $K=E$ and $\Theta> 0$ or if $K=J$ and $\sigma_{\mathbf{n}}\e_2>0$ (in which case $-i \Theta\geqslant 1$). On the other hand, if $K=J$ and $\sigma_{\mathbf{n}}\e_2<0$, then there remains a singularity at a distinct $t\in[0,1]$ for all the allowed values of $\Th$. Since both the pre-factor and the argument of the exponent blow up at this $t$-value, one may perform the integral by means of analytical continuation in the $t$-independent quantity in the exponent, resulting in that $\widehat S^\pm_{(K)}$ is analytic in $Y$ and $Z$ for any $x$ except at a proper subspace of twistor space. Whether this is an acceptable solution, and in particular whether there exists a gauge in which spacetime gauge fields can be extracted from it, remains to be examined. Finally, if $K=iB$ and $K=iP$ then $\Theta$ is indefinite and there is a finite region of spacetime in which $\Theta$ is real and hence $\widehat S^\pm_{(K)}$ is real-analytic in $Y$ and $Z$.

\item $\tilde L_{(K)}$ implements a different gauge choice on the deformed oscillators (and consequently on the gauge fields). The standard gauge choice of the perturbative expansion\footnote{Fixing this gauge leaves the (minimal) bosonic higher-spin algebra $\mhs_1(4)$ ($\mhs(4)$) as residual local symmetry algebra.} --- the aforementioned universal twistor gauge, $z^\a \wV_{\a} \ = \ 0$ --- is disrupted as soon as a non-trivial $\tilde L_{(K)}$ is introduced, since the latter rotates any $Z_{\underline\a}$ in the master-fields in a $K$-dependent way. While this does not affect any consideration based on HS-invariant quantities, it complicates the extraction of asymptotic charges --- especially in view of the fact that $\tilde L_{(K)}$ does not trivialize at the boundary of spacetime. Moreover, due to this gauge choice, a precise comparison of our solution based on ${\cal P}_1(E)$ with the Didenko-Vasiliev solution \cite{Didenko:2009td} still remains an open problem\footnote{In particular, while \eq{nsWeylspher} coincide with the Weyl 0-forms obtained in \cite{Didenko:2009td} for the case $n=1$ there treated, the deformed oscillators in \cite{Didenko:2009td} reduce to $Z_{\underline\a}$ asymptotically, while here (in the spherically-symmetric case) $\widehat S_{(K)\underline{\a}} \to\widehat S^{(0)}_{(K)\underline{\a}}= Z^{\tilde{L}_{(K)}}_{\underline{\a}}$ for $r\to \infty $ (see \eq{Sspher1}).}.  

\end{enumerate}

In the simplest spherically-symmetric case (\emph{i.e.} for the solution based on the ground-state projector ${\cal P}_1(E)$), the deformed oscillators can be written as 
\bea \widehat S^{\pm}_K& = & \tilde z^\pm+8\,{\cal P}_{1}(E^L)\,\tilde a^\pm\int^1_{-1}\frac{dt}{(t+1+i\s r(t-1))^2}\,j^\pm_{1}(t)\,e^{\,\ft{i\s( t-1)}{t+1+i\s r(t-1)}\,\tilde a^+\tilde a^- } \ ,\label{Sspher1}\eea
where $\tilde a^\pm:=\tilde u^{\a\pm} a_\a$ and $a_\a \ : = \ z_\a+i(\vark^L_\a{}^\b y_\b+v^L_\a{}^{\bd}\bar y_{\bd})$ coincide with the modified oscillators\footnote{Note that, due to the appearance of such modified oscillators, even choosing $\tilde L(x|Z)=1\!\!\!\!1$ the solution would not be in the universal twistor gauge. The additional complication that  $\tilde L(x|Z)$ introduces is that even the asymptotics of the spherically-symmetric solution are not in the universal twistor gauge (and reduce to $AdS_4$ only on the submanifold $Z=0$).} of \cite{Didenko:2009td}, obeying $z_\a\star {\cal P}_{1}(E)=a_\a  {\cal P}_{1}(E)$. Notice that, as anticipated, the internal connection for the spherically-symmetric case may only be non-analytic in $r=0$, as the form of the Weyl tensors \eq{nsWeylspher} suggests.

Once the deformed oscillators have been obtained, the generating functions of the gauge fields follows by computing their $\star$-product as in \eq{Wmu}. As shown in \cite{us}, since the $V^{\mathbf{n}}_\a$ self-replicate under $\star$-product, the gauge fields inherit the spacetime behaviour from the deformed oscillators, and are therefore regular for generic spacetime points.

We refer the reader to \cite{us} for the deformed oscillators in the general case and for the explicit form of the gauge fields generating functions in the general spherically-symmetric case.

\paragraph{\emph{Classical observables}.} 

In order to provide a gauge-invariant characterization of exact solutions that remains valid in strong-coupling regions where the weak-field expansion breaks down, it is useful to develop a formalism for classical observables.
These are functionals of the locally-defined master fields and transition functions, that are defined globally in generalized spacetimes carrying various higher-spin geometric structures \cite{Sezgin:2011hq}. 
There are several globally-defined formulations, or phases, of the theory, based on different unbroken gauge groups, or structure groups.
In what follows, we shall mainly focus on zero-form charges, which are observables in the unbroken phase\footnote{See \cite{us,Sezgin:2011hq} for certain $p$-form charges that may play an important r\^ole in the characterization of solutions.} and do not break any gauge symmetries. 
The basic such observables are Wilson loops in commuting sub-manifolds of ${\cal X}$.
These loops can be decorated with insertions of zero-form composites that transform as adjoint elements \cite{Sezgin:2011hq}. 
In the case of trivial monodromy, these can be contracted down to a single point ${\cal X}$ resulting in zero-form charges given by the generating function
\be {\cal I}(\s,k,\bar k;\l,\bar \l)~=~ {\widehat{\rm  Tr}}_{\Real}\left[(\widehat\k\widehat{\bar \k})^{\star\s}\star \exp_\star(\l^\a\widehat S_\a+\bar\l^{\ad}\widehat{\bar S}_{\ad})\star (\widehat \Phi\star\widehat\k)^{\star k}\star (\widehat \Phi\star\widehat{\bar \k})^{\star\bar k}\right]\ ,\ee
where $\widehat{\rm  Tr}_{\Real}$ is the chiral trace defined by 
\be \widehat{\rm Tr}_{\Real}[\widehat {\cal O}(Y,Z)]~=~ \int_{{\cal R}} {d^4Y d^4Z\over (2\pi)^4} \widehat {\cal O}(Y,Z)\ ,\label{chiraltrace}\ee
with $(y,z)$ and $(\bar y,\bar z)$ treated as real and independent variables; $(\s,k,\bar k)$ are natural numbers defined modulo $(\s,k,\bar k)\sim (\s\pm 2,k,\bar k)\sim (\s,k\pm 2,\bar k\mp2)\sim (\s\pm1,k\pm1,\bar k\mp1)$\,; and $(\l^\a,\bar\l^{\ad})$ are commuting spinors. 
The zero-form charges are manifestly higher-spin gauge invariant and hence defined globally on any base manifold; it follows that 
\be d\,{\cal I}(\s,k,\bar k;\l,\bar\l)~=~0\ ,\ee
modulo the equations of motion. The trace operation that defines them is also cyclic and independent of ordering prescriptions modulo possible boundary terms in twistor space. 
In what follows, we shall mainly be concerned with ${\cal I}(\s,k,\bar k):={\cal I}(\s,k,\bar k;0,0)$, 
and in particular with the supertraces 
\be {\cal I}_{2N}~:=~{\cal I}(1,2N,0)~=~\widehat{\rm Tr}_{\Real}[\widehat\kappa\widehat{\bar\kappa}\star (\widehat \Phi\star\pi(\widehat\Phi))^{\star N}]\ . \label{fullamp}\ee

Inserting the general expression of the Weyl zero-form \eq{generalWeyl0f} and using that $\k_y\star\k_y=1$ and the orthogonality and idempotency of the projectors, one gets
\bea {\cal I}_{2N}&:=& \widehat Tr_{\Real}[(\widehat\Phi\star\pi(\widehat\Phi))^{\star N}\star\widehat\kappa\widehat{\bar\kappa}]  \ = \ \widehat Tr_{\Real}[(\widehat\Phi'\star\pi(\widehat\Phi'))^{\star N}\star\widehat\kappa\widehat{\bar\kappa}]  \ = \ \sum_{\mathbf{n}\in (\integ +\ft12)^2}\nu_{\mathbf{n}}^{2N}\left.P_{\mathbf{n}}\right|_{Y=0}\ ,\label{chargesprime} \eea
for the axisymmetric projectors and analogously, substituting the double index $\mathbf{n}$ with the single index $n=\pm1,\pm2,...$  everywhere, for the symmetry-enhanced projectors ${\cal P}_n$. From the forms \eq{nonintproj} and \eq{enhanced2} (equivalently \eq{intproj} and  \eq{enhanced})  of the projectors it thus follows that
\bea {\cal I}_{2N}(K_{(+)},K_{(-)}) \ = \  4\sum_{\mathbf{n}\in (\integ +\ft12)^2}(-1)^{n_1+n_2-1} \nu_{\mathbf{n}}^{2N}  \label{0finvo2} \ ,\eea
for the axisymmetric solutions based on a given Cartan pair $(K_{(+)},K_{(-)})$, and
\bea {\cal I}_{2N}(K_{(q)}) \ = \  4\sum_{n=\pm1,\pm2,...}(-1)^{n-1}|n| \nu_{n}^{2N} \ ,\label{0finvenh}\eea
for the symmetry-enhanced ones, where we recall that the relation between $(n_1,n_2)$ and $n$ is $n:=q n_1+n_2$. 

Thus, the zero-form invariants ${\cal I}_{2N}$ extract, in general, a linear combination of powers of the deformation parameters $\nu_{\mathbf{n}}$ that characterize every solution, and that can be thought of as the eigenvalues of the expansion of the solution on the basis of projectors. For solutions based on a single projector (such as, for example, the BPS solution of \cite{Didenko:2009td}), these local invariants capture (even powers of) the unique deformation parameter sitting in front of the spin-two Weyl tensor as well as of its higher and lower-spin partners,  formally resembling the ADM mass. 

Interestingly, \eq{0finvenh} is \emph{not} divergent for any choice of (finite) deformation parameters, at least as long as the examined solution is based on finitely many projectors. This means that, for instance, although the rotationally-invariant Weyl curvatures \eq{nsWeylspher} asymptotically resemble those of a collection of ``higher-spin Schwarzschild black holes'', the apparent singularity in $r=0$ (\emph{i.e.}, in the strong-curvature region, where the pure spin-$2$ curvature invariants are no longer good observables) of the individual Weyl tensors does not actually lead to divergent higher-spin invariant zero-form charges\footnote{The regularity of these observables in $r=0$ may be traced back to the formal insensitiveness of the trace operation to the choice of ordering prescription. Indeed, at $r=0$ the Weyl-ordered symbol of this master-field is a distribution in twistor space \cite{us}. However, by moving to normal-ordering the resulting symbol becomes a regular, gaussian function. In this sense, the spacetime singularities may be resolved at the level of master-fields living in correspondence space.}. 

We defer the interesting issues of the physical significance and evaluation of other relevant invariants to a future publication.


\scs{Conclusions}


In this paper we have reviewed some properties of six infinite families of exact solutions to Vasiliev's four-dimensional higher-spin field equations, as well as the method through which they have been obtained. The latter is a combination of the gauge-function method, previously used for other exact solutions \cite{Sezgin:2005pv,Iazeolla:2007wt,Didenko:2006zd}, with an internal Ansatz generalizing that of \cite{Didenko:2009td}, based on the separation of the dependence of the master-fields on $Y$ and $Z$ twistor variables. The resulting solutions are organized in three pairs, each pair characterized by a biaxial isometry group $\mso(2)\oplus\mso(2)$ embedded into $\msp(4;\Comp)$ in three inequivalent ways. One of the families contains a subset of solutions in which one of the two $\mso(2)$ enhances to $\mso(3)$, while in the remaining families the enhanced symmetry algebra is $\mso(2,1)$. In all of our solutions, all spins are activated for generic choices of deformation parameters.

The study of the non-perturbative regime of higher-spin gravity is nowadays of extreme interest both in its own right, for uncovering the physics of higher-spin fields, and for its relevance in testing the proposed holographic duality \cite{Sezgin:2002rt}. In this sense, it is especially interesting to understand the extent to which the singular family ${\cal M}_{E}(E,J)$ can be thought of as a higher-spin generalization of black holes. To this purpose, it is crucial to carry out a more detailed study of whether, for instance, their singularities are physical and not a gauge artefact, and whether these solutions possess an event horizon. The latter question should in principle be addressed by analyzing the propagation of small fluctuations over them. 
The answers may have important surprises in store, since the deviations from Einstein gravity in the strong-curvature region, as discussed above, may be radical, essentially due to the non-locality of interactions induced by the unbroken higher-spin symmetry. 
To probe this region, it may be necessary to extend the usual tools of differential geometry to the higher-spin context, since standard concepts such as the relativistic interval are not higher-spin invariant. 

We have also reviewed the evaluation of certain zero-form charges \cite{Sezgin:2005pv,Iazeolla:2007wt,Colombo:2010fu,Sezgin:2011hq} on our solutions. They are a set of functionals of the zero-form master-fields, defined via the trace of the $\star$-product algebra, that are conserved on the field equations and provide useful instruments for distinguishing gauge-inequivalent solutions and for characterizing them physically even in strong-field regions. As the non-locality on ${\cal T}$ of the star-product is mapped via the field equations to spacetime non-locality, the zero-form charges hide their higher-derivative nature into the $\star$-products between master-fields, and this facilitates their evaluation. We find that certain zero-form charges involving the spacetime curvatures are well-defined on our solutions, and amount to linear combinations of powers of the squared deformation parameters $\m^2_{n}$, that therefore characterize the various field configurations in a gauge-independent way. Interestingly enough, all these invariants are finite everywhere (unless the solution under consideration is based on infinitely many projectors and the eigenvalues $\m_{n}$ are not too small). 

The study of exact solutions in higher-spin gravity also prompts us to tackle some crucial issues that have remained poorly explored so far: a global description of the solutions (and correspondingly adapted choices of gauge functions), the characterization of boundary conditions and superselection sectors within the framework of the unfolded equations formulated on correspondence spaces (see \cite{us} for more details), the extraction of asymptotic charges and symmetries, and the description of possible global degrees of freedom carried by the $Z$-dependence are but a few. 

Finally, it would also be interesting to extend the Ansatz here presented to Kerr-like solutions and to study whether or not it is possible to generalize it to the construction of multi-soliton configurations.


\vspace{1.5cm}

{\bf Acknowledgments}

We are grateful to M.~Ammon, G.~Barnich,  N.~Boulanger, A.~Castro, D.~Chialva, N.~Colombo, D.~Fioravanti, D.~Francia, M.~Gaberdiel, S.~Giombi, D.~Grumiller, E.~Joung, P.~Kraus, C.~Maccaferri, E.~Perlmutter, J.~Raeymaekers, A.~Sagnotti, M.~Schnabl, M.~Taronna, A.~Waldron, X. Yin and especially to V.~E.~Didenko and M.~A.~Vasiliev  for stimulating discussions. C.~I. gratefully acknowledges the ``Universit\`a degli Studi `G. Marconi' '' in Rome for partial support. 

\vspace{1cm}



\end{document}